\documentclass[hidelinks,12pt]{article}
\usepackage{hyperref}
\usepackage{setspace,amsmath,amssymb,amsthm,mathrsfs,sectsty,xcolor,graphicx,geometry,authblk,rotating,lscape}
\usepackage[square,sort,comma,numbers]{natbib}
\usepackage{epstopdf}
\usepackage[normalem]{ulem}
\useunder{\uline}{\ul}{}
\hypersetup{
colorlinks,
linkcolor={red!50!black},
citecolor={red!50!black},
urlcolor={blue!50!green}
}
\allsectionsfont{\centering}
\begin{document}

\title{Exponential Self-Organization and Moore's Law: Measures and Mechanisms}

\author[1]{\small{Georgi Georgiev}\thanks{ggeorgie@assumption.edu; georgi@alumni.tufts.edu; ggeorgiev@wpi.edu (Corresponding Author)}}
\author[2]{Atanu Chatterjee\thanks{atanu@wpi.edu}}
\author[3]{Germano Iannacchione\thanks{gsiannac@wpi.edu}}
\affil[1]{Department of Physics, Assumption College, 500 Salisbury St., Worcester, MA, 01609, USA} 
\affil[1]{Department of Physics, Tufts University, 4 Colby St., Medford, MA, 02155, USA}
\affil[1,2,3]{Department of Physics, Worcester Polytechnic Institute, 100 Institute Road, Worcester, MA, 01609, USA}

\date{}

\maketitle

\begin{abstract}
\noindent
The question how complex systems become more organized and efficient with time is open. Examples are, the formation of elementary particles from pure energy, the formation of atoms from particles, the formation of stars and galaxies, the formation of molecules from atoms, of organisms, and of the society. In this sequence, order appears inside complex systems and randomness (entropy) is expelled to their surroundings. Key features of self-organizing systems are that they are open and they are far away from equilibrium, with increasing energy flowing through them. This work searches for global measures of such self-organizing systems, that are predictable and do not depend on the substrate of the system studied. Our results will help to understand the existence of complex systems and mechanisms of self-organization. In part we also provide insights, in this work, about the underlying physical essence of the Moore's law and the multiple logistic growth observed in technological progress.
\end{abstract}


\section{Introduction}
Important questions in contemporary physics remain unanswered: Why and how complex systems self-organize? How does this process occur in accordance with the Second Law of Thermodynamics? What are the relationships and interactions between the different characteristics of complex systems that make them function and lead to the decrease of their internal entropy? The answers to these and other related questions are urgent and crucial, since numerous phenomena in various fields of science are dependent on them. Chemistry needs to explain how autocatalytic cycles form and change with time to improve their efficiency. Biology needs to understand how an organism\rq{}s metabolism becomes more efficient in using energy and time for their functioning. Economics needs to explain the increase of efficiency of different technologies and networks in society. The complexification of systems over time has been a subject of considerable scientific interest for years. It has been noted that as systems grow they become more intricate and complex as can be seen in everything around us, from stars and galaxies to forests and cities ~\citep{bonner2004perspective, chaisson2004complexity, chaisson2002cosmic, west1997general, bettencourt2007growth}. 

Why do we need to know the mechanism of self-organization? How will this help us? First of all, science has always been driven by the quest to understand unexplained phenomena. As soon as we understand them, we can use them for our benefit. We live in one such self-organizing system, our society, and we ourselves, as biological organisms are self-organized entities as well. Therefore to explain how complex systems function and self-organize further, is of utmost importance. Without explanatory power, we do not have the ability to understand and improve the systems that we live in. In the field of complex systems the process of progressive development is understood as a continuous improvement through self-organization. New structures, rules and laws in systems emerge at the new levels of organization. But, how is organization defined, and how it and the rate of self-organization are to be measured and quantified? What quantitative measures can be used to describe them? What are the mechanisms, the potential for further self-improvement in complex systems and their limits? The answers to those questions are vital and will help us understand more deeply physical, chemical, biological and economic complex systems. 

To answer the above questions, we apply a new measure to quantify organization complexity and the rate of self-organization based on the Principle of Least Action~\citep{euler1952methodus, lagrange1853mecanique, de1744accord, de1746loix}. The fundamental nature of this principle allows all the conservation laws and equations of motion, in all branches of physics, from Classical Physics to General Relativity and Quantum Mechanics, to be derived from it.

The quantity action ($A$), is given as the integral of a system\rq{}s Lagrangian over time where the Lagrangian is the difference between the kinetic and potential energies at each instant along a path or a trajectory, written as: 

\begin{equation}
A = \int_{t_1}^{t_2} (T-V)dt = \int_{t_1}^{t_2} \mathcal{L}(p_{i},q_{i})dt
\end{equation}

\noindent Here $T$ denotes the Kinetic Energy, $V$ the potential energy, the pair $(q_{i},p_i)$ is the position and momentum vector in the generalized coordinates, and $\mathcal{L}(\cdot)$ the Lagrangian functional. The Hamiltonian formalism of the Action Principle imposes constraints on the end-points (say, A and B) of the trajectory, such that both end-points and end-times, $(A, t_A)$ and $(B, t_B)$ are known, and $A(t_{A},A)=A(t_{B},B)=0$. This makes the problem completely deterministic, whereas in nature most often, the fate of a particle (or a system of particles) is completely unknown. Although, Maupertuis' formulation of the Action Principle removes the time constraints, yet it still requires the end-points of a path to be defined. According to Maupertuis' formulation, the action can be given by: 

\begin{equation}
A=\int_{q_1}^{q_2} p_{i}dq_{i}=\int_{q_1}^{q_2} 2\sqrt{T_i}dq_{i}=\int_{t_1}^{t_2} 2T_{i}dt
\end{equation}

\noindent where, $p_{i}=\frac{\partial \mathcal{L}}{\partial \dot{q_i}}$. Processes in nature occur only when the action is minimized (Hamilton\rq{}s formulation) or along those trajectories that minimize action (Maupertuis\rq{} formulation), i.e., $\delta A = 0$. It is important to understand that the minimization of action, the product of energy and time or position and momentum, is central and not the minimization of energy and time separately. A simple thought experiment (\textit{gedankenexperiment}) will reveal that minimization of energy and time separately either do not yield any self-organization or they are forbidden by the existing laws of physics. Minimizing energy ($\delta E \sim 0$) yields an equilibrium state for a system, such as a crystal without any flows of energy or changes in entropy and consequently no change in the current state of the self-organization. Similarly, minimizing time ($\delta t \sim 0$), results in violating the relativistic limit of the speed of light. Even if we imposed such a limit, the amount of energy necessary for the motion increases to infinity, therefore maximizing the action. Therefore, a balance must exist between energy and time for natural processes and the minimization is not of the two individual entities separately, but of their product, the action. 

Since, complex systems undergoing self-organization are open systems, far away from equilibrium, energy and matter pass through them along the paths of least obstructive constraint, they can be represented as flow networks formed by those paths. The nodes in these systems act as sources and sinks, and edges as trajectories, along which the system elements flow. These elements are prevented from moving along their least action paths by the presence of obstructive constraints within the system. The total action of the system is the sum of all individual actions for all agents and all edge crossings per unit time, $\Sigma_{i,j} A_{ij}$, where the indices `$i$\rq{} and `$j$\rq{} represent the $i^{th}$ agent\rq{}s $j^{th}$ edge crossing. The smallest unit of action is one quantum of action, which is an universal constant denoted by \lq{}$h$\rq{} (the Planck\rq{}s constant) where $n$ is an integer:

\begin{equation}
A = \int_{t_1}^{t_2} E_i dt = \int_{q_1}^{q_2} p_i dq_i = nh
\end{equation}

The sequential motion of elements in a complex system from a source to a sink is one edge-crossing. We define this as an event in space-time. Therefore, organization $\alpha$, as the action efficiency, is the ratio of the number of events occurring in a system to the total amount of action in a given interval of time. Multiplication by the Planck's constant makes the measure dimensionless and defines it as reciprocal to the total number of quanta of action per event. The less the average action per event, the closer the system is to the attractor state \emph{i.e.}, the least action state. Action efficiency as a measure for the amount of organization in a system, is inversely proportional to the average action per event. It is time dependent as it changes in self-organization and evolutionary processes:

\begin{equation}
\alpha = \frac{nmh}{\Sigma_{i, j} A_{i, j}}
\end{equation}

\noindent where the indices, `$i$\rq{} and `$j$\rq{} sum upto the integers `$n$\rq{} and `$m$\rq{} respectively, where $n$ is the number of elements, and $m$ is the number of edge crossings per element per unit time. Just as water is diverted by rocks as it flows down the stream, the system\rq{}s elements are sometimes blocked from traversing along the path of least time and energy by the presence of obstacles. With self-organization, the system elements perform work on the obstructive constraints and minimize them. We model the time dependence of organization in these systems as increase in efficiency of physical action, where the action efficiency is defined as the decrease of action for an element of a system to traverse a pair of nodes along its flow networks. Therefore, the state of organization of a system can simply be described by the position of the constraints, and the minimization of the constraints in respect to the flows leads to an increased action efficiency per unit motion. This is analogous to the stream of water doing work on the rocks that block its most efficient path until the rocks are moved. This is expressed quantitatively in terms of energy and temporal efficiency of processes as observed in nature~\citep{georgiev2014mechanism}.

In previous papers, we defined least unit action to be the product of the least amount of time and energy needed to make a single edge crossing in a flow network~\citep{georgiev2014mechanism, georgiev2002least, georgiev2012quantitative}. The flow ($\phi$) in a system is defined as the total number of edge crossings by the elements (agents) per unit of time in the system. Representing organized systems as flow networks implies a constant flow of energy and matter which by definition means the system must be far from equilibrium. These systems are open, with branching hierarchical networks and fractal like self-similar structures~\citep{smyth2003conductivity}. The characteristics of these natural systems include fractal like properties that change and grow in a manner which have a universal predictability. We see evidence of this in natural systems everywhere, from cardiovascular networks to cities~\citep{bettencourt2007growth, rozenfeld2009fractal}. In many such systems the scaling laws, which are generally power-law relations, $f(y)\sim y^{\delta}$, define the scale-free properties of change, with $\delta$ being the scaling exponent~\citep{goh2001universal, barabasi1999emergence, barrat2008dynamical, chatterjee2015statistical}. Due to the presence of scaling relationships and power-law decays in the statistical properties of the networks, the importance of the respective nodes is non-uniform. Certain nodes have been found to be relatively more important, `central\rq{}, as compared to the others. In order to capture the relative importance of the various nodes in the networks various centrality measures are calculated, such as degree, betweenness and closeness~\citep{kitsak2007betweenness, albert2002statistical}. In many real world networks the property of self-similarity has been found to be of significant interest as in many systems, the system elements overcome jamming, which is an obstructive constraint to their motion (which decrease flow, lowering action efficiency and ultimately organization), by branching out and forming self-similar patterns~\citep{tang2009self, blagus2012self, zhang2007adaptive, chatterjee2015studies}.

In our earlier work, we have shown that decreasing the unit action between two nodes will increase the overall action efficiency of a sample system, the core processing unit of computers. The CPU data were collected from Intel Corporation~\footnote{http://www.intel.com} in order to solve for the smallest amount of action per computation, as well as the total amount of action, within a certain time interval. The results of the data analysis showed that the organization (quality) and total action (quantity) both increased exponentially over time which is in agreement with the quality-quantity relationships noted in the literature ~\citep{bonner2004perspective, georgiev2014mechanism, carneiro1967relationship}. The data also demonstrated that this relationship between quantity and quality is a power law, which matched well with the predictions of our model. The data showed that the least unit action of the CPU\rq{}s ($\alpha$) and the total amount of action ($Q$) are in a positive feedback loop, leading to an exponential growth of both and power law relationship between the two. As the efficiency of the total action increases, more time and energy are freed to further reorganize the system and decrease unit action. A system with high action efficiency allows the system to grow in quantity of action. Therefore the total amount of action was found to be in a positive feedback with the organization of the system, as more time and energy are necessary to achieve further constraint minimization and action efficiency. 

According to our previous papers, the Principle of Least Action explains the mechanism of increase of organization through quantity accumulation and constraint and curvature minimization with an attractor, the least average sum of actions of all elements and for all motions. In this study, we present more measures to quantify self-organization in complex systems. We also develop a mathematical model to capture the presence of positive feedback loops between these measures. This is necessary in order to, first, understand all the links between all the measures leading to the causal change in each of them, and second, to increase the amount of information that we can gather from complex systems, in order to quantify the process of self-organization in them. When these measures show increase in time, they do so according to the power law proportionality, and because of the positive feedback connections between them, they are all inter-dependent functions of each other. Establishing links and proportionality relationships, in the form of power laws between them, will help us to calculate the values for some of these measures in systems where these measures are hard to obtain directly once we know the rest of the quantities. 

Those characteristics of complex systems are mutually interdependent, because one of them can increase only if the rest have increased to a certain level. This interdependence allows us to call those functions participating in a circular positive feedback mechanism – interfunctions. Those interfunctions increase together and can deviate from their proportionality values, when the system is in dynamic equilibrium, which can be called homeostasis. The interfunctions can deviate from those homeostatic values by a certain amount, beyond which the mechanism of interaction between them is disturbed. Therefore negative feedback exists to restore their homeostatic values, which is proportional to the difference between their actual values and their homeostatic values. The homeostatic level increases exponentially due to the positive feedback between them and the actual values oscillate around the exponential homeostatic value due to the restoring force of their deviations. Thus, the system of interfunctions acts as a system of coupled harmonic oscillators around their exponentially growing homeostatic values. In real world systems, those values are perturbed by random external noise. Thus, the fluctuations of the system of harmonic oscillators become stochastic. The modeling and analysis of those oscillations around the exponential trends, observed in the data in this paper, will be an object of further work. It can explain the origin of multiple logistic growth observed in technology substitution curves. 

We broaden the system of interfunctions including the action efficiency and total amount of action in a complex system, based on a system of ordinary differential equations: (\textbf{i}) which leads to exponential growth with time, and (\textbf{ii}) establishes a power relation between the two, with measures such as the total flow of events, which is the number of computations for the CPUs and the number of transistors. Our study can also explain the origin of the observed exponential change in technology, noticed empirically by Moore~\cite{moore}, Kurzweil~\cite{kurzweil}, Nagy et al.~\cite{nagy}, and Kelly~\cite{kelly}. This understanding can help describe, quantify, measure, manage, design and predict future behavior of complex systems to achieve the highest rates of self-organization to improve their quality. Our long-term goal is to test whether it can be applied to complex systems across disciplines not only from Physics, but also from Chemistry, Biology, Ecology, and Economics. 

\section{Theory: Correlation of Quantities}
The variational approach to describe systems in nature is becoming increasingly important~\citep{pernu2012natural, hartonen2012natural, makela2010natural, chatterjee2012action, chatterjee2013principle, chatterjee2015thermodynamics}. The results presented in this paper are a continuation of the variational approach that was previously used to show that minimizing unit action is correlated to maximizing total action in increasing the level of organization of complex systems~\citep{georgiev2014mechanism}. In this paper we study the correlation between efficiency of unit action (action efficiency) $\alpha$ and total action $Q$ with transistor count $N$ and flow of events, which for CPUs are computations, $\phi$. The goal is to establish a connection between flow and physical size as characteristic measures for self-organization in physical systems, and correlate them with action efficiency. For biological systems and processes it is challenging to measure the physical quantity of action. Therefore, to expand the applicability of our theory, and to make it more widely usable for the scientific community, we hope to find other correlated characteristic measures that can be used to derive the more fundamental ones. By exploring these correlated characteristics that are more accessible, we can calculate the unit action from the already available quantities that participate in the positive feedback loop. This will allow us to study larger varieties of systems and to generalize our present theory to systems of any nature across all disciplines. 

\begin{figure}[t]
\begin{center}
\includegraphics[width=0.6\linewidth]{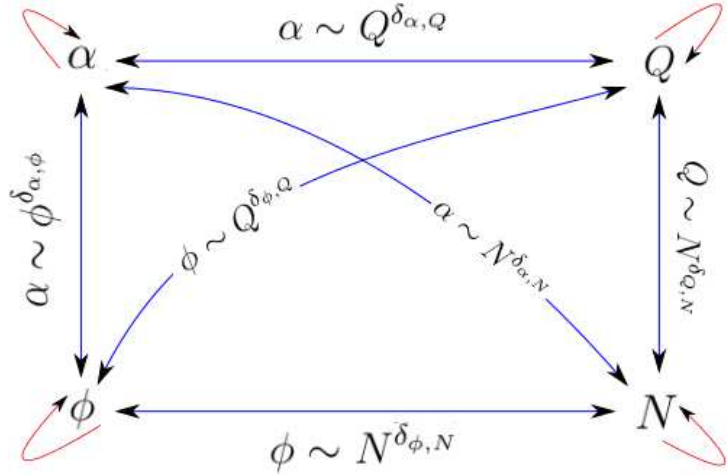}
\caption{Figure shows the positive feedback loop among the system variables, $\alpha,\phi,Q,N$ and their corresponding scaling relationships.}
\end{center}
\end{figure}

The following system of equations was observed as a relation between each of the four variables, $\alpha, \phi, Q, N$. The time derivatives of these variables are denoted as $\dot{\alpha}$, $\dot{\phi}$, $\dot{Q}$, $\dot{N}$. Since, there exist a positive feedback loop between these variables, the governing set of differential equations is written as: 

\begin{equation}
\left.
\begin{aligned}
\dot{\alpha} &= a_{11}\alpha + a_{12}\phi + a_{13}Q + a_{14}N\\
\dot{\phi} &= a_{21}\alpha + a_{22}\phi + a_{23}Q + a_{24}N\\
\dot{Q} &= a_{31}\alpha + a_{32}\phi + a_{33}Q + a_{34}N\\
\dot{N} &= a_{41}\alpha + a_{42}\phi + a_{43}Q + a_{44}N
\end{aligned}
\right\}
\end{equation}

The above system of equations can be written in a compact form for a function, $f(t,\alpha,\phi,Q,N)$ with $a_{ij}$ being the coefficients as,

\begin{equation}
\dot{f}_i(t,\alpha,\phi,Q,N) = \Sigma_{i,j=1}^4 a_{ij}f_j (t,\alpha,\phi,Q,N)
\end{equation}

Since, the only independent variable is the time, $t$, the above system of equations can be represented in the matrix form as,
\begin{equation}
\dot{\textbf{F}}(t) = \mathcal{\textbf{A}} \textbf{F}(t)
\end{equation}
The coefficients $a_{ij}$ belong to the matrix $\textbf{A}$, and the vector, $\textbf{F}(t)$ takes on the characteristics, $\alpha,\phi,Q,N$ for various values of the indices $i$ and $j$ for all $i, j\in \{1,2,3,4\}$. The solution to the above equation is given by,
\begin{equation}
f_i (t) = \Sigma_{i=1}^{4} c_i\exp(\lambda_i t)\mathcal{\textbf{u}}_i
\end{equation}
where $\lambda_i$ is the Eigenvalue, $\mathcal{\textbf{u}}_i$ is the respective Eigenvector, and $c_i$ an arbitrary constant, at time, $t = 0$. Since, the system variables are in a positive feedback loop (see above and Figure 1), the coefficients, $a_{ij}$ are positive for all $i$, $j$. The standard solution to Equation 7 will yield both positive and negative Eigenvalues (in Equation 8). At steady-state, $t\rightarrow\infty$, $\exp(-\lambda_i t)\rightarrow 0$ and $\exp(\lambda_i t)$ only persists. Thus, the solution for the system variables with respect to time can be written as,
\begin{equation}
\left.
\begin{aligned}
\alpha\sim\exp(\lambda_{\alpha}t)\\
\phi\sim\exp(\lambda_{\phi}t)\\
Q\sim\exp(\lambda_{Q}t)\\
N\sim\exp(\lambda_{N}t)
\end{aligned}
\right\}
\end{equation}
where $\lambda_{\alpha}$, $\lambda_{\phi}$, $\lambda_{Q}$ and $\lambda_{N}$ are the exponential scaling exponents for the system variables. Interestingly, upon elimination of the independent variable, $t$ (or setting $a_{ii}=a_{jj}=0$), power-law relationships (or scale-free relationships) can be established between the system variables with respect to each other, which for the current theme of our study is very important. Therefore, Equation 9 can be rewritten as, 

\begin{equation}
\left.
\begin{aligned}
\alpha\sim \phi^{(\lambda_{\alpha}/\lambda_{\phi})}\sim Q^{(\lambda_{\alpha}/\lambda_{Q})}\sim N^{(\lambda_{\alpha}/\lambda_{N})}\\
\phi\sim Q^{(\lambda_{\phi}/\lambda_{Q})}\sim N^{(\lambda_{\phi}/\lambda_{N})}\\
Q\sim N^{(\lambda_{Q}/\lambda_{N})}\\
\end{aligned}
\right\}
\end{equation}
We introduce a new scaling parameter, $\delta$ and rewrite the set of Equation 10 as,
\begin{equation}
\left.
\begin{aligned}
\alpha\sim \phi^{\delta_{\alpha, \phi}}\sim Q^{\delta_{\alpha, Q}}\sim N^{\delta_{\alpha, N}}\\
\phi\sim Q^{\delta_{\phi, Q}}\sim N^{\delta_{\phi, N}}\\
Q\sim N^{\delta_{Q, N}}\\
\end{aligned}
\right\}
\end{equation}

\section{Data and Methods}
The data was collected from the Intel Corporation Datasheets for the CPUs from 1971 to 2013. The Instructions Per Second (IPS) for each processor was divided by the Thermal Design Power (TDP) as a measure of the total power consumption by the CPUs at maximum computational speed, for consistency. The result was multiplied by the table value of the Planck\rq{}s constant $h=6.626\times 10^{-34} Js$, as the smallest quantum of action, to solve for $\alpha$, as the inverse of the number of quanta of action per one instruction per second. The TDP was divided by the Planck\rq{}s constant, $h$ to find the total number of quanta of action per second, $Q$. Only processors for desktops or laptops were used, because some of the specialized processors, such as the ones for phones or tablets, perform slower in order to consume less energy.

\section{Results and Discussion}

\begin{table}[t]
\centering
\begin{tabular}{|c|l|cl|c|l|c|l|}
\hline
\multicolumn{8}{|c|}{{\ul \textbf{Exponential scaling exponents}}} \\ \hline
\multicolumn{2}{|c|}{{\color[HTML]{3531FF} \textbf{$\lambda_{\alpha}$}}} & \multicolumn{2}{c|}{{\color[HTML]{3531FF} \textbf{$\lambda_{\phi}$}}} & \multicolumn{2}{c|}{{\color[HTML]{3531FF} \textbf{$\lambda_Q$}}} & \multicolumn{2}{c|}{{\color[HTML]{3531FF} \textbf{$\lambda_N$}}} \\ \hline
\multicolumn{2}{|c|}{$7.76\times 10^{-9}$} & \multicolumn{2}{|c|}{$1.24\times 10^{-8}$} & \multicolumn{2}{c|}{$4.71\times 10^{-9}$} & \multicolumn{2}{c|}{$1.05\times 10^{-8}$} \\ \hline
\multicolumn{8}{|c|}{{\ul \textbf{Power-law scaling exponents}}} \\ \hline
\multicolumn{2}{|c|}{{\color[HTML]{329A9D} \textbf{$\delta_{\alpha, \phi}$}}} & \multicolumn{2}{c|}{{\color[HTML]{329A9D} \textbf{$\delta_{\alpha, Q}$}}} & \multicolumn{2}{c|}{{\color[HTML]{329A9D} \textbf{$\delta_{\alpha, N}$}}} & \multicolumn{2}{c|}{{\color[HTML]{329A9D} \textbf{$\delta_{\phi, Q}$}}} \\ \hline
\multicolumn{2}{|c|}{0.61} & \multicolumn{2}{c|}{1.39} & \multicolumn{2}{c|}{0.72} & \multicolumn{2}{c|}{2.39} \\ \hline
\multicolumn{4}{|c|}{{\color[HTML]{329A9D} \textbf{$\delta_{\phi, N}$}}} & \multicolumn{4}{c|}{{\color[HTML]{329A9D} \textbf{$\delta_{Q, N}$}}} \\ \hline
\multicolumn{4}{|c|}{1.17} & \multicolumn{4}{c|}{0.45} \\ \hline
\end{tabular}
\caption{Table outlines the various scaling exponents that were derived in Equations 9 and 11 and their respective magnitudes as calculated from the data in Figures 2 and 3.}
\end{table}

\begin{figure}[hb!]
\begin{center}
\includegraphics[width=0.9\linewidth]{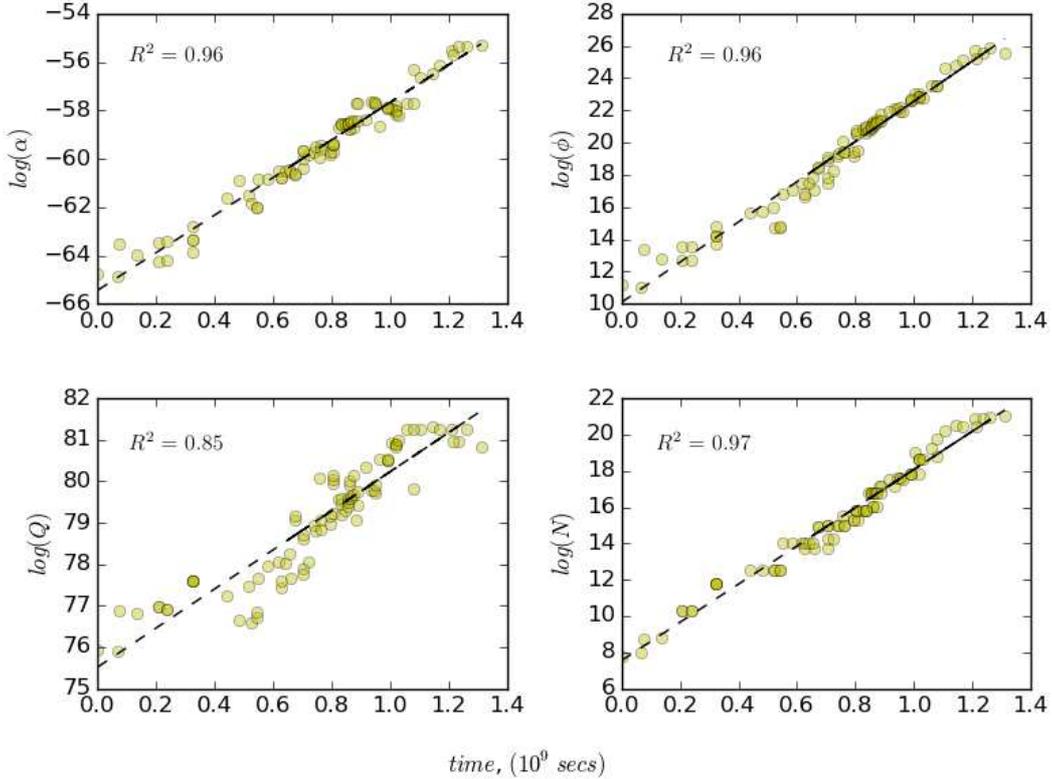}
\caption{Figure shows the exponential scaling relationships between the characteristics, $\alpha, \phi, Q$ and $N$ with respect to time on a semi-logarithmic scale (see Equation 9) with the goodness of fit (inset).}
\end{center}
\end{figure}

In Figure 1 we represented graphically the interdependence between the interfunctions. In Figure 2, we show that the interfunctions, $\alpha, \phi, Q$ and $N$ obey an exponential relationship with respect to time. The exponential scaling exponents, $\lambda$ are tabulated in Table 1. We plot the dependence of each system variable against the other in Figure 3. On eliminating time from the system of exponential scaling equations (Equation 9), we observe that the system variables relate to each other through power-laws. The power-law exponents, $\delta$ are also tabulated in the table above (see Table 1). The scale-free exponents in Table 1 arise out from the mutual permutations among the system parameters, as can be easily observed from Equations 10 and 11. We define the proportionality constants for the system of equations connecting $\alpha$, $\phi$, $Q$ and $N$ with time, $t$ as follows:

\begin{equation}
\left.
\begin{aligned}
\alpha = C_{\alpha}\exp(\lambda_{\alpha}t)\\
\phi = C_{\phi}\exp(\lambda_{\phi}t)\\
Q = C_{Q}\exp(\lambda_{Q}t)\\
N = C_{N}\exp(\lambda_{N}t)
\end{aligned}
\right\}
\end{equation}

\noindent The terms $C_{\alpha}$, $C_{\phi}$, $C_{Q}$ and $C_{N}$ are the proportionality constants for the system of exponential relations as represented in Equation 9. Similarly, the power-law relationships between $\alpha$, $\phi$, $Q$ and $N$ (see Equation 11) with the proportionality constants $C_{\alpha, \phi}$, $C_{\alpha, Q}$, $C _{\alpha, N}$, $C_{\phi, Q}$, $C_{\phi, N}$ and $C_{Q, N}$ can be rewritten as:

\begin{equation}
\left.
\begin{aligned}
\alpha = C_{\alpha, \phi} \phi^{\delta_{\alpha, \phi}}\\
\alpha = C_{\alpha, Q} Q^{\delta_{\alpha, Q}}\\
\alpha = C _{\alpha, N} N^{\delta_{\alpha, N}}\\
\phi = C_{\phi, Q} Q^{\delta_{\phi, Q}}\\
\phi = C_{\phi, N} N^{\delta_{\phi, N}}\\
Q = C_{Q, N} N^{\delta_{Q, N}}\\
\end{aligned}
\right\}
\end{equation}

We tabulate the constants from the plots (Figure 2 and 3) in Table 2. The matrix elements can be obtained from the proportionality constants among the interfunctions. The element, $a_{ij} = C_{ij}$, $a_{ji} = a_{ij}^{-1/\delta_{ij}}$, and $a_{ii}=a_{jj}=0$ for all $i,j$ where the indices run over the functions $\alpha,\phi,Q,N$. 

\begin{table}[hb]
\centering
\begin{tabular}{|c|l|cl|c|l|c|l|}
\hline
\multicolumn{8}{|c|}{{\ul \textbf{Proportionality Constants}}} \\ \hline
\multicolumn{2}{|c|}{{\color[HTML]{3531FF} \textbf{$C_{\alpha}$}}} & \multicolumn{2}{c|}{{\color[HTML]{3531FF} \textbf{$C_{\phi}$}}} & \multicolumn{2}{c|}{{\color[HTML]{3531FF} \textbf{$C_Q$}}} & \multicolumn{2}{c|}{{\color[HTML]{3531FF} \textbf{$C_N$}}} \\ \hline
\multicolumn{2}{|c|}{$4\times 10^{-29}$} & \multicolumn{2}{|c|}{$23920$} & \multicolumn{2}{c|}{$6\times 10^{32}$} & \multicolumn{2}{c|}{$1878$} \\ \hline

\multicolumn{2}{|c|}{{\color[HTML]{329A9D} \textbf{$C_{\alpha, \phi}$}}} & \multicolumn{2}{c|}{{\color[HTML]{329A9D} \textbf{$C_{\alpha, Q}$}}} & \multicolumn{2}{c|}{{\color[HTML]{329A9D} \textbf{$C_{\alpha, N}$}}} & \multicolumn{2}{c|}{{\color[HTML]{329A9D} \textbf{$C_{\phi, Q}$}}} \\ \hline
\multicolumn{2}{|c|}{$9\times 10^{-32}$} & \multicolumn{2}{c|}{$2\times 10^{-74}$} & \multicolumn{2}{c|}{$2\times 10^{-31}$} & \multicolumn{2}{c|}{$2\times 10^{-74}$} \\ \hline
\multicolumn{4}{|c|}{{\color[HTML]{329A9D} \textbf{$C_{\phi, N}$}}} & \multicolumn{4}{c|}{{\color[HTML]{329A9D} \textbf{$C_{Q, N}$}}} \\ \hline
\multicolumn{4}{|c|}{3.84} & \multicolumn{4}{c|}{$2\times 10^{31}$} \\ \hline
\end{tabular}
\caption{Table outlines the various proportionality constants for the system of Equations 9 and 11 as rewritten in Equation 12 and 13.}
\end{table}

We represent the matrix elements $a_{ij}$ for the interfunctions in a compact manner as follows:
\begin{equation}
\begin{pmatrix}0&9\times 10^{-32}&2\times 10^{-74}&2\times 10^{-31}\\7.84\times 10^{50}&0&2\times 10^{-74}&3.84\\1.04\times 10^{53}&6.86\times 10^{30}&0&2\times 10^{31}\\4.33\times 10^{42}&0.31&2.76\times 10^{-70}&0\\\end{pmatrix}
\end{equation}

In order to understand how the interfunctions vary with respect to each other we plot the scaling relationships in Figure 4. It is interesting to observe that action efficiency, $\alpha$ and total action, $Q$ scale much faster than a linear relationship. This observation illuminates the question, how physical systems progressively self-organize with time. It means that action efficiency needs to increase faster than the quantity of the system, in order to accommodate the increased amount of action in it. It does that by  increasing the flow through it, for example, by developing flow channels, tributaries, veins and vesicles. Hence, the flow, $\phi$, is always observed to be super-linear against all of the interfunctions. It means that the flow of events is always ahead of any of the increase of any of the other interfunctions. At the other extreme, the total action, $Q$, trails the increase of any other interfunction.

\begin{figure}[hb]
\begin{center}
\includegraphics[width=0.9\linewidth]{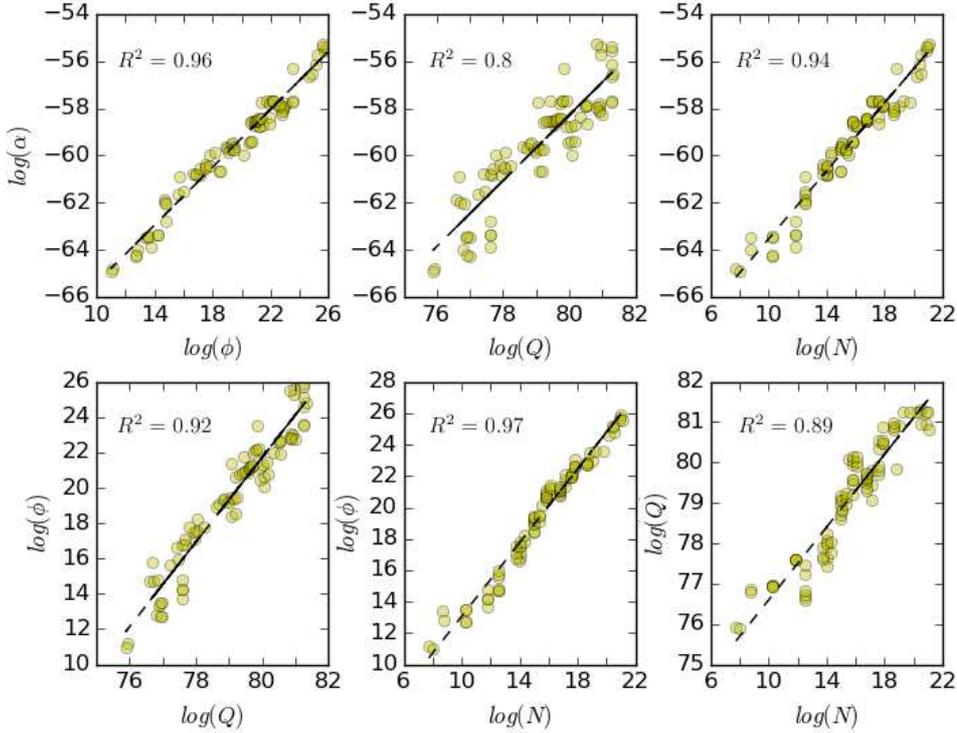}
\caption{Figure shows the power-law scaling relationships between the characteristics, $\alpha, \phi, Q$ and $N$ on a double-logarithmic scale (see Equation 11) with the goodness of fit (inset).}
\end{center}

\end{figure} 
\begin{sidewaysfigure}
\begin{center}
\includegraphics[width=1\linewidth]{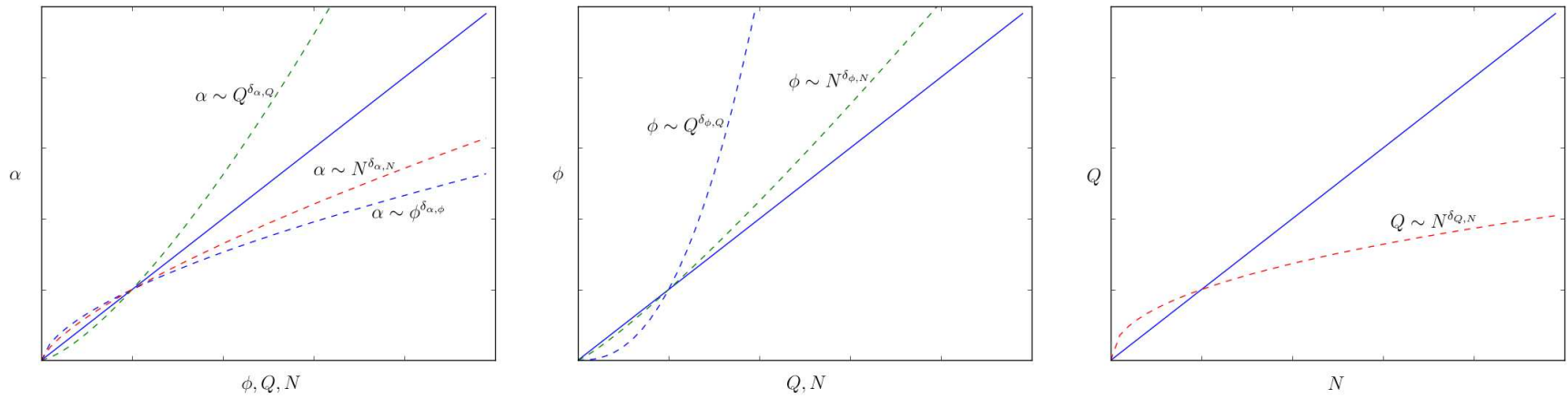}
\caption{Figure presents a pictorial representation of the power-law scaling relationships between the system variables, $\alpha, \phi, Q$ and $N$, and the solid diagonal line signifies slope of one.}
\end{center}
\end{sidewaysfigure}

From the data shown, we can see evidence that flow of events, $\phi$, which is the number of computations per second, and total number of transistors $N$ grow exponentially in time and are in a power law relation of each other, similar to organization, $\alpha$ and total amount of action, $Q$. In previous work we showed that when elements organize to achieve least unit action per unit event in the system, the overall capacity for flow of the events in the network increases. By looking at our new results, we observe that $\phi$ and $N$ can both be used as quantitative characteristics that are in a positive feedback relation with $\alpha$ and $Q$. This means that we can potentially use $N$, and flow as proxies to measure quality and quantity increase in self-organization of complex systems when those are hard to measure directly. The flow in biological and social self-organizing systems, such as the flow of events in metabolic cycles, which are chemical reactions, or social transactions, is often much more accessible than action and action efficiency, which are the more fundamental quantities. Analogous to the number of transistors in the CPUs, we can use as a quantitative characteristic the number of cells in an organism or the number of people in a city or a society.

The Moore's law and the other observations of exponential change in technology, are a part of this model. Moore's law empirically describes the exponential increase of the number of transistors in time, which agrees with the solutions of our model of a positive feedback between the number of transistors and other interfunctions, such as action efficiency, flow of events, and the total amount of action. Thus, the observation by Moore is explained here as a part of this system of interfunctions driven by the Principle of Least Action. The visible oscillations of the data around the exponential and power law fits (see Figure 2 and 3), which are their homeostatic values, can help explain the multiple logistic nature of technology substitution $S$--curves, by the negative feedback between the homeostatic values of the interfunctions, and the actual deviations of the data from them.  

\section{Conclusions}
The results have shown that flow of events, $\phi$ and number of transistors, $N$, for CPUs are in a positive feedback with the action efficiency, $\alpha$, and the total action $Q$ and also with each other. The positive feedback between them forms a system of coupled differential equations with steady state solutions indicating exponential growth in time. An important prediction of the model is that these mutually dependent characteristics of a self-organizing system - the interfunctions - are in power law relationship with each other. Therefore, all four quantities are proportional to each other at different stages of organization and we can use any one of them to solve for the other three. The predictions of our model fit well with the data, and they also provide an insight to understand the physical nature of technological progress as observed by Moore and others using the fundamental concepts of time and energy. Therefore, this opens before us an opportunity, to learn about the processes in those self-organizing dynamic systems, by measuring just one or a few of those interfunctions and deduce the others. This is a significant opportunity to study a variety of systems, because in some self-organizing systems certain interfunctions are more accessible than others, and the most fundamental, of action efficiency and total amount of action are the hardest to obtain. Hence, it is crucial to find further characteristic measures in a self-organizing complex system, as it will allow us to study a wide variety of these systems under the same framework even with partial information. An important follow up work is to compare the coefficients in the equations for this system with other self-organizing systems in nature and look for universal constants. The origin of the oscillations around the homeostatic values in the data, may illuminate multiple logistic curves in technology, such as the substitution $S$--curves. \\ \\

\section*{Acknowledgments}
\textbf{G G} thanks Assumption College for financial support and encouragement of this research through faculty development, sabbatical and course load reduction grants, and for financial support for undergraduate research. He also thanks Worcester Polytechnic Institute for his affiliation there and for support for graduate student research. \textbf{A C} and \textbf{G I} thank the Department of Physics at Worcester Polytechnic Institute for providing the resources to carry out the study. All the authors thank the reviewers for their insightful comments and constructive suggestions. 
\vspace{0.08in}

\noindent The authors declare that there is no conflict of interest regarding the publication of this paper.

\begin{center}
\rule{10cm}{1pt}
\end{center}

\end{document}